\documentclass[prm, preprint, longbibliography]{revtex4-1}
\usepackage{graphicx}
\usepackage{subfigure}
\usepackage{amsmath}
\usepackage{amsfonts}
\usepackage{empheq}
\graphicspath{{./figures/}}
\usepackage[bookmarks, hyperindex, colorlinks=true,
            linkcolor=red,
            urlcolor=blue, citecolor=blue]{hyperref}
\usepackage{listings}
\usepackage{braket}
\usepackage{bm}
\usepackage{url}
\renewcommand{\vec}{\bm}

\usepackage{color}
\makeatletter
\begin{document}
\title{Dielectric dependent hybrid functionals for heterogeneous materials}
\author{Huihuo Zheng}
\email{huihuo.zheng@anl.gov}
\affiliation{Argonne Leadership Computing Facility, Argonne National Laboratory, Lemont, IL 60439, USA}
\author{Marco Govoni}
\email{mgovoni@anl.gov}
\affiliation{Materials Science Division, Argonne National Laboratory, Lemont, IL 60439, USA}
\affiliation{Institute for Molecular Engineering, University of Chicago, Chicago, IL 60637, USA}
\author{Giulia Galli}
\email{gagalli@uchicago.edu}
\affiliation{Materials Science Division, Argonne National Laboratory, Lemont, IL 60439, USA}
\affiliation{Institute for Molecular Engineering and Department of Chemistry, University of Chicago, Chicago, IL 60637, USA}
\begin{abstract} 
We derive a dielectric-dependent hybrid functional which accurately describes the electronic properties of heterogeneous interfaces and surfaces, as well as those of three- and two-dimensional bulk solids. The functional, which does not contain any adjustable parameter, is a generalization of self-consistent hybrid functionals introduced for homogeneous solids, where the screened Coulomb interaction is defined using a spatially varying, local dielectric function. The latter is determined self-consistently using density functional calculations in finite electric fields.  We present results for the band gaps and dielectric constants of 3D and 2D bulk materials, and band offsets for interfaces, showing an accuracy comparable to that of GW calculations.
\end{abstract}
\date{\today}
\maketitle

\section{Introduction}
Density Functional Theory (DFT) was first applied to compute the structural and electronic properties of condensed systems more than 35 years ago \cite{Yin1982, Parr1995, Martin2004}, using the local density approximation \cite{Ceperley1980, Perdew1981} of the exchange and correlation (xc) energy functional. Approximately ten years later, when gradient corrected approximations (GGA) \cite{Langreth1980, Langreth1983, Hu1985, Perdew1992, Perdew1996, Zhang1998} for the xc energy were derived, DFT was adopted for some molecular investigations by the quantum chemistry community. Shortly after the first GGA molecular calculations, hybrid functionals were proposed \cite{Becke1993_new_mixing, Becke1993_exact_exchange, Becke1996, Perdew1996_PBE0, Jaramillo2003, Chai2008, Maier2018} and most DFT applications for finite systems, which use localized basis sets, have been carried out with hybrid functionals \cite{BAUSCHLICHER1995, Ren2012}, most notably B3LYP \cite{Lee1988, Becke1993_exact_exchange, Stephens1994}. These are functionals where the exchange energy is defined as a linear combination of exact (Hartree-Fock) and local exchange \cite{Ghosh2018}. The condensed matter physics community adopted hybrid functionals later than the quantum chemistry community, due to computational difficulties in evaluating the Hartree-Fock (HF) exchange energy using plane wave (PW) basis sets; these are the basis set of choice in most of the codes used for materials \cite{ABINIT2002, ABINIT2009, Gygi2008, QE-2009, VASPurl, CP2Kurl, CPMDurl, Lejaeghere2016}, although periodic DFT codes using localized basis sets are also in use \cite{SIESTA2001, CRYSTAL2018, g16, Blum2009}.  The difficulties in evaluating HF exchange in PW basis sets have now been largely overcome, with the advent of fast algorithms based on bisection techniques \cite{Gygi2009, Gygi2013, Dawson2015, Lin2016} or maximally localized Wannier functions \cite{Wu2009, DiStasio2014}. Nevertheless periodic DFT calculations with hybrid functionals and PW basis sets remain substantially heavier, from a computational standpoint, than local or semi-local DFT calculations. The functionals PBE0 \cite{Adamo1999} and HSE \cite{Heyd2003, Heyd2004, Krukau2006} are among the most popular hybrid functionals used for condensed systems, and lately dielectric dependent hybrid functionals \cite{Skone2014,Skone2016,Brawand2017,Gerosa2018_Accuracy} have been increasingly used to predict structural and electronic properties of solids \cite{Skone2014,Ferrari2015, Skone2016,Seo2016, PhysRevMaterials.2.073803, Chen2018, Gerosa2015, Gerosa2018, Gerosa2018_Accuracy, Kronik2018,doi:10.1021/acs.jctc.7b01058} and liquid \cite{Gaiduk2016,Pham2017,Gaiduk2018} and of several molecules \cite{Skone2016,Brawand2016,Brawand2017}. Another category of orbital dependent functionals recently proposed is that of Koopmans-compliant functionals, used for both molecules and solids \cite{PhysRevLett.105.266802,Borghi2014,Nguyen2018}. 

A drawback of most of the functionals  mentioned above is that while they work well for certain classes of homogeneous systems, e.g. solids, they are usually not as accurate for heterogeneous systems, e.g. surfaces and interfaces, where the dielectric screening of different portions of the system differ substantially. For heterogeneous semiconductors, Shimazaki \textit{et al.} \cite{Shimazaki2015} introduced an estimator of the electrostatic environment surrounding the atoms in a semiconductor leading to the definition of position-dependent atomic dielectric constants. For solid/solid interfaces, Borlido \textit{et al}. \cite{Borlido2018} introduced a non-local mixing fraction, based on an estimator of a local dielectric function that contains parameters to be evaluated with system-dependent fitting procedures.

In this work, we propose a hybrid functional that describes equally accurately three- and two-dimensional solids, as well as surfaces and interfaces, and which is derived entirely from first principles, with no need to define any adjustable parameter. The functional is based on an approximation of the screened Coulomb interaction using a local dielectric function, which is derived from first principles by minimizing a dielectric enthalpy functional.  We first discuss (Section \ref{sec:disentanglement}) the foundation of dielectric disentanglement by showing that the dielectric screening of a system composed of two subsystems interfaced with each other, may be decomposed into the screening of the two subsystems plus an interfacial contribution. The disentanglement is carried out using a localized representation of the eigenvectors of the dielectric matrix, obtained using bisection techniques originally proposed for the eigenfunctions of Kohn-Sham Hamiltonians \cite{Gygi2009}. Our results on dielectric decomposition are used to justify the definition of a local, spatial dependent dielectric function (Section \ref{sec:disentanglement}), which in the bulk portion of the subsystems coincides with their respective dielectric constants.  We then use this local dielectric function to define a dielectric hybrid functional for heterogeneous systems (Section \ref{sec:schyb}); the functional is derived from first principles, without any adjustable parameter, by carrying out calculations in finite electric field. Finally we present results for 3D and 2D solids in Section \ref{sec:bulk} and for surfaces and interfaces in Section~\ref{sec:surfaces_interfaces}, with focus on the calculations of band gaps, dielectric constants and band-offsets.

\section{Spatial disentanglement of dielectric spectra}
\label{sec:disentanglement}
In this section, we address the following question: can the dielectric matrix of an heterogeneous system (composed, e.g. of two solids or a liquid and a solid) be expressed in terms of the dielectric matrices of the subsystems? For simplicity we restrict our attention to a system of volume $\Omega$ composed of two subsystems, A and B interfaced with each other and we consider a single interface between A and B. We address the question by writing a spectral decomposition of the dielectric matrices of the heterogeneous system and of A and B, and then  we use bisection techniques \cite{Gygi2009} to localize the eigenvectors of the dielectric matrices in desired regions of space. 

According to linear response theory,  the density-density response function $\chi$  and the irreducible polarizability $\chi^0$
are related to the dielectric matrix ($\epsilon$) of the system by the following equation: 
\begin{eqnarray}\label{eq:1}
\bar \epsilon &=& 1 - \bar \chi^0, \quad \bar \chi = \frac{\bar \chi^0}{1-\bar \chi^0}\,,
\end{eqnarray}
where the bar in Eq.~\eqref{eq:1} indicates that the functions have been symmetrized with respect to the Coulomb potential (see, e.g. Ref. \cite{Govoni2015}).
We represent $\bar \chi^0$ using its spectral decomposition, 
\begin{eqnarray}
\bar{\chi}^0(\vec r, \vec r') = \sum_n \lambda_n \phi_n(\vec r)\phi^*_n(\vec r')\,,
\end{eqnarray}
where $\phi_n$ and $\lambda_n$ are eigenvectors and eigenvalues, respectively. In the following we focus on static dielectric responses. 

\begin{figure*}[hbt]
\centering
\includegraphics[width=0.90\linewidth]{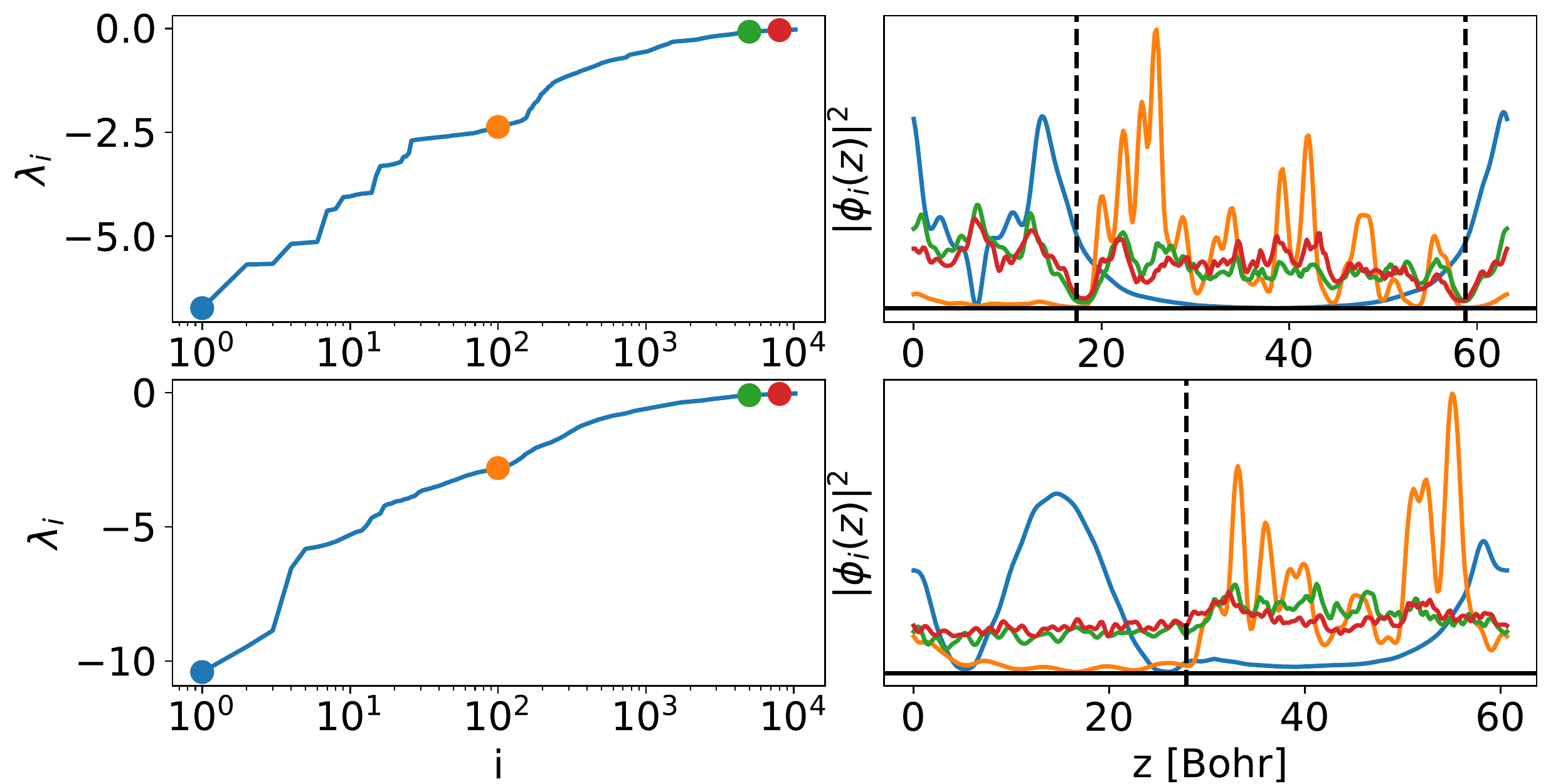}
\caption{Spectral decomposition of the response function $\bar \chi^0$ of H-Si/H$_2$O (upper panels) and Si/Si$_3$N$_4$ (lower panels) interfaces.  Left: eigenvalues of $\bar\chi^0$; the points correspond to eigenvectors shown on the right panel. Right: selected eigenpotentials (labeled by different colors) projected on the axis (z) perpendicular to the interface: $|\phi_i(z)|^2 = \frac{1}{L_x L_y}\int dxdy |\phi_i(x, y, z)|^2$. The black vertical dashed lines denote the position of the interface and are determined according to the spatial variation of the charge density (see Fig.~\ref{fig:bipartition})}\label{fig:sp_dcp}
\end{figure*}

Fig.~\ref{fig:sp_dcp}  shows the eigenvalues of $\bar \chi^0$ (left panels) for two representative interfaces,  H-Si/H$_2$O and Si/Si$_3$N$_4$, one where the two subsystems are non-covalently bonded and one where there are covalent bonds at the interface (the geometry of the model slabs and how they were obtained are described in the SM \cite{SM}). The square moduli of selected eigenvectors projected in the direction perpendicular to the interface ($z$) and their corresponding eigenvalues (dots) are shown on the right and left panels of Fig.~\ref{fig:sp_dcp}, respectively.  For both surfaces, we see that some eigenmodes are predominantly localized on one side of the slabs while other modes, especially those corresponding to $|\lambda_i| \to 0$ (green and red curves) are localized over the entire slab.  

In order to express response functions of the entire system in terms of those of the subsystems, we represent the dielectric matrix in terms of localized functions, instead of eigenfunctions. 
\begin{figure*}[hbt]
\centering
\includegraphics[width=0.90\textwidth]{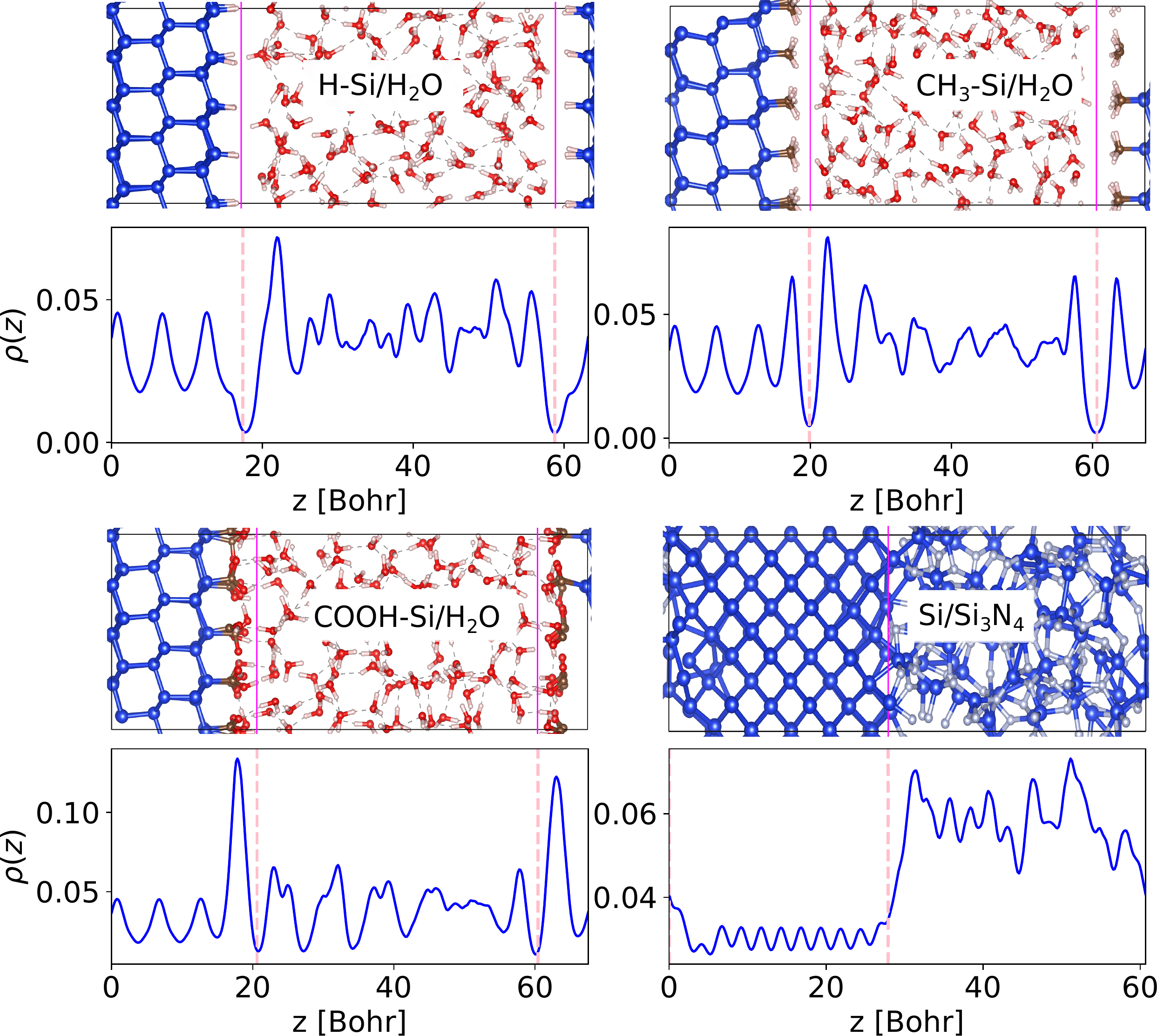}
\caption{Charge density $\rho$ (e/Bohr$^3$) of four slabs representing solid/liquid and solid/solid interfaces projected on the axis z perpendicular to the interface [$\rho(z) = \frac{1}{L_xL_y}\int dx dy dz \rho(\vec r)$]. Vertical red lines represent the position of the interfaces and were determined based on the spatial variation of the charge density. }\label{fig:bipartition}
\end{figure*}
We first define two subsystems using the projection of the electronic charge density on the $z$-axis perpendicular to the interface, as illustrated in Fig.~\ref{fig:bipartition}. The use of the charge density to define regions A and B introduces a certain degree of arbitrariness, as a criterion is required to determine charge density minima, in correspondence of which interface planes are defined.  Such a criterion is system dependent. While  the charge density is used in this section to define interfacial planes for the purpose of illustrating the concept of disentanglement of the dielectric response, it will not be used in practical calculations. As we will see in Section~\ref{sec:finite_field},  a general, system independent procedure can be defined to compute local dielectric functions.   

After partitioning the full system into subsystems using the charge density, we obtain a set of localized functions from the set of eigenvectors $\phi_i$ by constructing and diagonalizing the filtered overlap matrix $\mathcal{M}$,  
\begin{eqnarray}
&& \mathcal{M}_{ij} := \int_{\vec r \in \Omega_S} d\vec r \phi^*_i(\vec r) \phi_j (\vec r),\nonumber\\&&\mathcal{M}\cdot V_m = w_m V_m, \quad w_m\in [0, 1]\,,
\end{eqnarray}
where $w_m$ and $V_m$ are eigenvalues and eigenvectors of  $\mathcal{M}$, and $\Omega_S$ is the volume of either subsystem A or B as defined using the electronic charge density (see Fig.~\ref{fig:bipartition}). 
The set of eigenvectors of $\mathcal{M}$ provides the transformation matrix from the set of $\phi_i(\vec r)$'s to a set of localized orbitals.
The eigenvalues $w_m$ represent the weights of the localized orbital $\phi^\text{loc}_m(\vec r)$ within the subspace $\Omega_S$: 
\begin{eqnarray}\label{eq:weight}
w_m =\frac{\int_{\vec r \in \Omega_S} d\vec r |\phi_m^\text{loc}(\vec r)|^2}{\int d\vec r |\phi^\text{loc}_m(\vec r)|^2}\,.
\end{eqnarray}
If $w_i\simeq 1$, $\phi_i^\text{loc}$ is localized on $\Omega_S$; if $w_i\simeq 0$, $\phi_i^\text{loc}$ is localized on $\Omega-\Omega_S$. We classify the  $\phi_i^\text{loc}(\vec r)$'s into three subsets: 
\begin{eqnarray}\label{eq:basis_sets}
\mathcal{F}_A &=&\Big \{\phi_i^\text{loc}\Big|w_i<w_{thr}\Big\}, \nonumber\\
\mathcal{F}_B &=& \Big\{\phi_i^\text{loc}\Big|w_i>1-w_{thr}\Big\}, \nonumber\\
\mathcal{F}_I &=& \Big\{\phi_i^\text{loc}\Big|w_{thr}<w_i<1-w_{thr}\Big\}\,,
\end{eqnarray}
where $w_{thr}$ is a chosen localization threshold that can be systematically varied to verify the robustness and convergence of the localization procedure (it was chosen to be 0.01 in the examples shown in the figures). 

\begin{figure*}[hbt]
\centering
\includegraphics[width=0.90\linewidth]{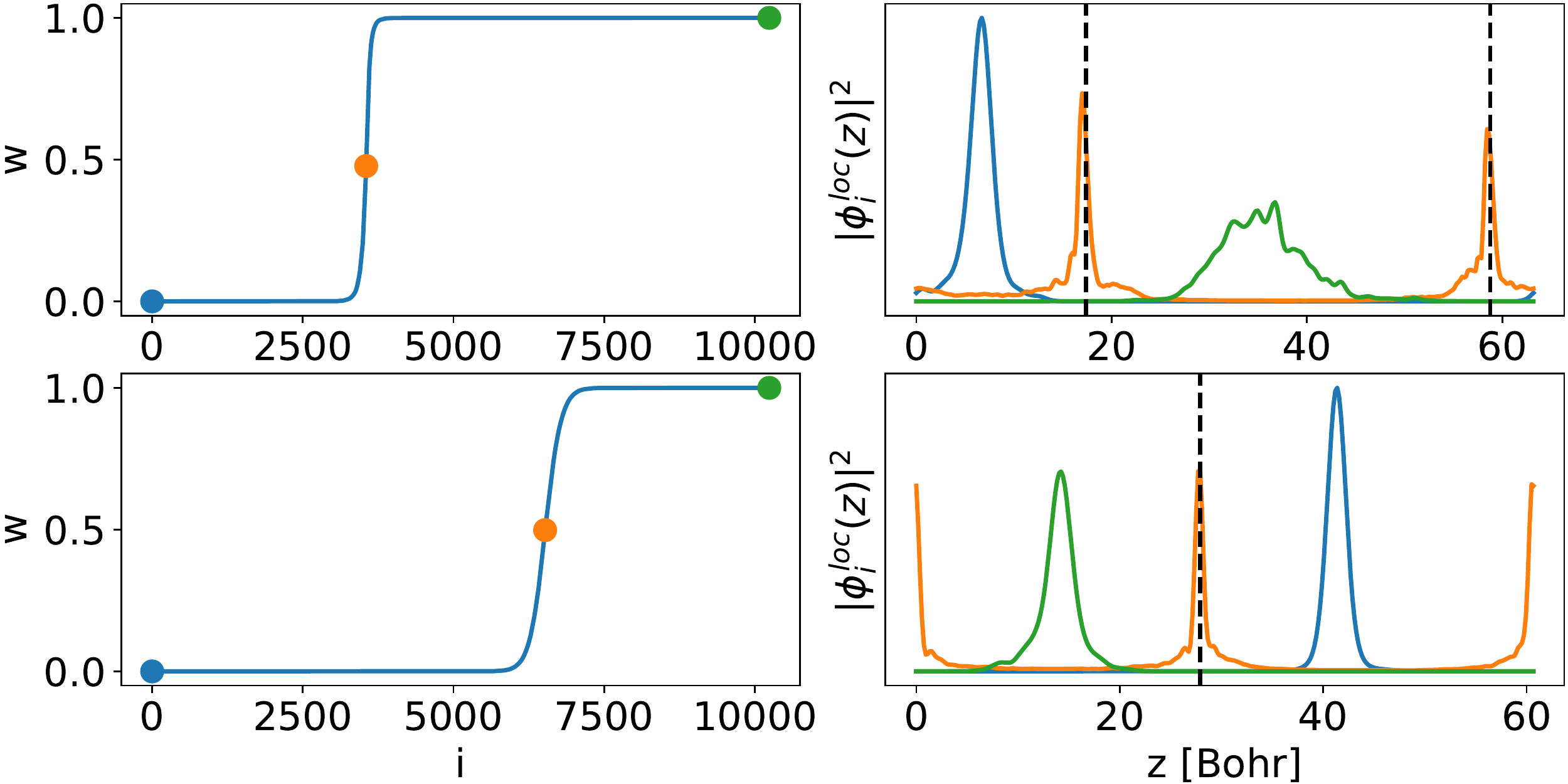}
\caption{Weights (w, left panels) of bisected localized potentials for two interfaces, as defined in Eq.~\eqref{eq:weight}, and representative bisected localized potentials ($\phi^{loc}$, right panels), projected on the direction $z$ perpendicular to the interface. The localized potentials have been obtained from the eigenpotentials  of $\bar \chi_0$ for the H-Si/H$_2$O (upper panels) and Si/Si$_3$N$_4$ (lower panels) interfaces. The dots on the left panels correspond to the localized potentials shown on the right panels.  In our calculations we included 10,240 eigenpotentials in the spectral decomposition of the irreducible polarizability and we verified that such number yielded a converged results for the localized orbitals and weights shown in the figure.}\label{fig:bis_loc}
\end{figure*}

Fig.~\ref{fig:bis_loc} displays the weights $w$ and the square moduli of localized basis functions for the H-Si/H$_2$O and Si/Si$_3$N$_4$ interfaces: we found that most of the basis functions are localized in one of the two subsystems, with the rest of them localized near the interface.

After obtaining the localized basis set $\mathcal{F}$ ($\mathcal{F}_A \cup \mathcal{F}_B \cup \mathcal{F}_I $), we expressed the matrix elements of $\bar{\chi}^0$ as 
$\bar \chi^0 = \bar \chi^0_A + \bar \chi^0_B + \bar \chi^0_I + \bar \chi^0_\text{off-diag.}$, where $\chi^0_\text{off-diag.}$ includes all the off-diagonal blocks representing the coupling between the two subsystems. 
By diagonalizing $\bar \chi^0_A$, $\bar \chi^0_B$ and $\bar \chi^0_I$ in the respective subspaces $\mathcal{F}_A$, $\mathcal{F}_B$, and $\mathcal{F}_I$ defined in Eq.~\eqref{eq:basis_sets}, we found that the response of the whole system can be disentangled into contributions from the subsystems, i.e. we found that for all systems studied here: 
\newcommand{\eig}{\text{eig}}
\begin{eqnarray}\label{eq:decompose}
\eig(\bar{\chi}^0_A) \cup \eig(\bar{\chi}^0_B) \cup \eig(\bar{\chi}^0_I) \simeq \eig(\bar{\chi}^0)\,.
\end{eqnarray}

\begin{figure*}[hbt]
\centering
\includegraphics[width=0.90\textwidth]{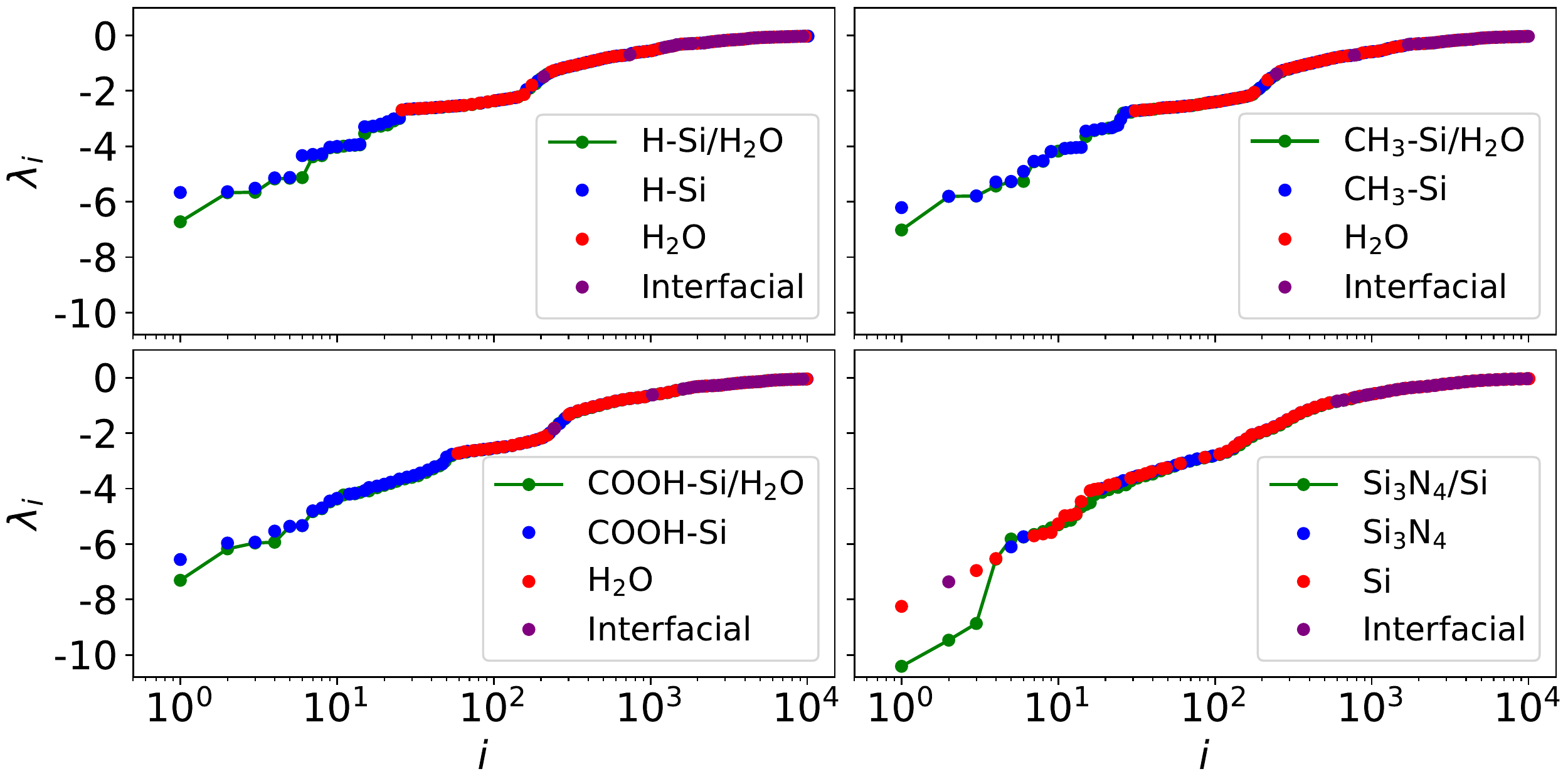}
\caption{Disentanglement of the dielectric spectra of several interfacial systems. The eigenvalues ($\lambda$) of the subsystems (dots) are compared with those of the whole system (solid curve) to verify the validity of  Eq.~\eqref{eq:decompose}.}\label{fig:decompose_comp}
\end{figure*}

Fig.~\ref{fig:decompose_comp} shows decomposed spectra [Eq.~\eqref{eq:decompose}] compared with the spectrum of the whole system. It is seen that  [$\eig(\bar{\chi}^0_A) \cup \eig(\bar{\chi}^0_B) \cup \eig(\bar{\chi}^0_I)$] and $\eig(\bar{\chi}^0)$ give very similar results, with small differences in the low eigenvalue regions, due to the neglect of the elements of $\chi^0_\text{off-diag.}$. As expected neglecting these elements is a better approximation for aqueous interfaces than for the Si-Si$_3$N$_5$ interface, where covalent bonds are formed. Therefore, we conclude that the dielectric screening of the whole slab may be approximated as the sum of contributions from the subsystems plus an interfacial dielectric screening contribution.

The results of this section indicate that  it is reasonable to approximate the screening of the entire slab by a local dielectric function $\epsilon(\bf r)$, a smooth function expected to describe accurately the screening of the two separate subsystems in their respective bulk regions. We will see in the next section that these assumptions  lead to a definition of a generalized dielectric hybrid functional which yields accurate band gaps and dielectric constants for 2D and 3D systems and band offsets for complex interfaces. 

We now turn to describing a procedure to obtain $\epsilon(\bf r)$ which does not rely on the definition of an interface plane based on the electronic charge density, nor on any parameters defining subsystems A and B.

\section{Dielectric-dependent hybrid functionals}
\label{sec:schyb}
The  results on dielectric disentanglement described in the previous section led to the idea of defining a local dielectric function whose limiting values in the two subsystems is expected to coincide with the dielectric constants of the respective bulk subsystems. Such a local dielectric function can then be used to generalize the hybrid functionals introduced in Ref.~\cite{Skone2014}. In the following, we define the local dielectric function from first principles.

In Ref.~\cite{Skone2014}, the screened Coulomb interaction in a homogeneous system is approximated as $W(\vec r, \vec r') = \frac{1}{\epsilon_\infty |\vec r-\vec r'|}$, where $\epsilon_\infty$ is the macroscopic static dielectric constant. This approximation is used in the definition of a hybrid functional similar to PBE0 but with mixing fraction $\alpha = 1/\epsilon_\infty$ instead of 0.25.
Several authors have suggested using $\alpha$ as an adjustable parameter to reproduce the experimental band gap of solids \cite{PhysRevLett.101.106802,doi:10.1002/pssb.201046195,SHIMAZAKI200891,PhysRevB.83.035119,PhysRevB.88.081204,CONESA201311,Gerosa2018_Accuracy}.

Following the definition of exchange in Ref.~\cite{Skone2014, Borlido2018}, if we write the screened Coulomb potential as
\begin{eqnarray}
W(\vec r, \vec r') =\alpha(\vec r, \vec r')\frac{1}{|\vec r - \vec r'|}\,,
\end{eqnarray}
the exchange energy of the entire system takes the following form \cite{Borlido2018}
\begin{eqnarray}\label{eq:exc_hetero1}
E_{x} &=&-\sum_{i< j}\int d\vec r d\vec r' \alpha(\vec r, \vec r')\frac{\psi_i^*(\vec r)\psi_j^*(\vec r')\psi_j(\vec r)\psi_i(\vec r')}{|\vec r - \vec r'|} \nonumber\\
&&+ \int d\vec r \Big [1-\alpha(\vec r, \vec r)\Big ]\rho(\vec r)e_x^\text{PBE}[\rho(\vec r)]\,.
\end{eqnarray}
We assume that the function  $\alpha(\vec r, \vec r')$ is a simple separable function of $\epsilon(\vec r)$  and $\epsilon(\vec r')$, with  $\alpha(\vec r, \vec r)=\epsilon (\vec r)$ and we write:
\begin{eqnarray}
\alpha(\vec r, \vec r')\simeq\frac{1}{\sqrt{\epsilon(\vec r)  \epsilon(\vec r')}}\label{eq:separableform}  \,.
\end{eqnarray} 
We then arrive at the following ansatz for the exchange and correlation energy: 
\begin{eqnarray}\label{eq:exc_hetero2}
E_{xc} &=&-\sum_{i< j}\int d\vec r d\vec r' \frac{1}{\sqrt{\epsilon(\vec r) \epsilon(\vec r')}}\frac{\psi_i^*(\vec r)\psi_j^*(\vec r')\psi_j(\vec r)\psi_i(\vec r')}{|\vec r - \vec r'|} \nonumber\\
&&+ \int d\vec r \Big [1-\frac{1}{\epsilon(\vec r)}\Big ]\rho(\vec r)e_x^\text{PBE}[\rho(\vec r)] \nonumber\\
&& + \int d\vec r\rho(\vec r)e_c^\text{PBE}[\rho(\vec r)]\,. 
\end{eqnarray}
where we have chosen the PBE approximation to represent the local part of the exchange and correlation energy. 

The exchange-correlation functional defined in Eq.~\eqref{eq:exc_hetero2} is similar, in spirit, to the local functional proposed in Ref.~\cite{Borlido2018}. However we emphasize two important conceptual and practical differences:  we have provided a theoretical justification of Eq.~\eqref{eq:exc_hetero2} based on the decomposition of the screened Coulomb interaction into that of subsystems and an interfacial region. Next we show that $\epsilon(\vec r)$ may be obtained from first principles by carrying out calculations in finite field, eliminating the need to tune any arbitrary parameter, or adopt any fitting, system-dependent procedure, which are necessary instead in the formalism of Ref.~\cite{Borlido2018}.

\section{Self-consistent determination of local dielectric functions using a finite field approach}
\label{sec:finite_field}
Here we describe a finite field approach to compute $\epsilon (\vec r)$. In general, the macroscopic dielectric tensor of any condensed system can be obtained by carrying out calculations in a finite electric field and by minimizing the functional \cite{Umari2002, Souza2002, Stengel2009}: 
\begin{eqnarray}\label{eq:E}
F(\vec E, [\rho]) &=&E_\text{KS}[\rho] + \int V(\vec r)\rho(\vec r) d\vec r = E_\text{KS}[\rho] - \int \vec E \cdot \vec r \rho(\vec r) d\vec r\,, 
\end{eqnarray}
where $\int \vec E \cdot \vec r \rho(\vec r) d\vec r $ is called the electric enthalpy, and $E_\text{KS}$ is the Kohn-Sham energy of the system. Alternatively one could minimize the functional:  
\begin{eqnarray}\label{eq:D_functional}
&&U(\vec D, [\rho])= E_\text{KS}[\rho] + \frac{1}{8\pi}\int d\vec r (\vec D-4\pi \vec P)^2 \,,
\end{eqnarray}
where  $\vec D = \vec E + 4\pi\vec P = \vec \epsilon \cdot \vec E $, and 
$\vec P$ is the polarization of the system; the components of the dielectric tensor $\epsilon$ are:
\begin{eqnarray}\label{eq:eps_ij}
\epsilon_{\alpha\beta}=\delta_{\alpha\beta} + 4\pi \frac{\partial P_\alpha}{\partial E_\beta}, \quad (\epsilon^{-1})_{\alpha\beta}=\delta_{\alpha\beta} - 4\pi \frac{\partial P_\alpha}{\partial D_\beta}\,,
\end{eqnarray}
where $\alpha$ and $\beta$ are Cartesian coordinates. In periodic systems, the induced polarization can be computed from the shift of the centers of the Wannier functions ($\Delta \vec r^i_c$) of the unperturbed system when an electric field is applied \cite{King1993,Stengel2006}. For a homogeneous system of $N_\text{s}$ occupied states, the average change in macroscopic polarization is given by
\begin{eqnarray}\label{eq:polarization_const}
\Delta \vec P = \frac{-e}{\Omega} \sum_{i=1}^{N_{\text{s}}} \Delta \vec r^i_c \,.
\end{eqnarray}
 This allows us to define a spatial dependent polarization for heterogeneous systems (e.g., 2D materials, surfaces and interfaces):
\begin{eqnarray}\label{eq:polarization}
\Delta \vec P(\vec r) = -e\sum_{i=1}^{N_\text{c}} N_i \Delta \vec R_c^i  \delta (\vec r - \vec R_c^i)\,,
\end{eqnarray}
where $N_\text{s}$ Wannier centers have been grouped in $N_\text{c}$ clusters: $N_i$ is the number of Wannier centers in the $i$-th cluster, $\vec \Delta R_c^i = \frac{1}{N_i}\sum_{j=1}^{N_i} \vec \Delta r_c^j$ is the shift of the center of the $i$-th cluster induced by the applied electric field. In practical calculations the $\delta$-function is replaced by a Gaussian function of finite width equal to the average of the spreads of the corresponding Wannier orbitals belonging to the same cluster. We note that $\Delta \vec P$ entering Eq.~\eqref{eq:polarization_const} can be obtained from $\Delta \vec P(\vec r)$ using the following relation:
\begin{equation}
\Delta \vec P = \frac{1}{\Omega} \int_\Omega \Delta \vec P (\vec r ) d\vec r \,.
\end{equation}
The spatial dependence of $\epsilon$ is then defined by the spatial dependence of the polarization, as given in Eq.~\eqref{eq:polarization}.

\begin{figure}[hbt]
    \centering
    \includegraphics[width=0.50\textwidth]{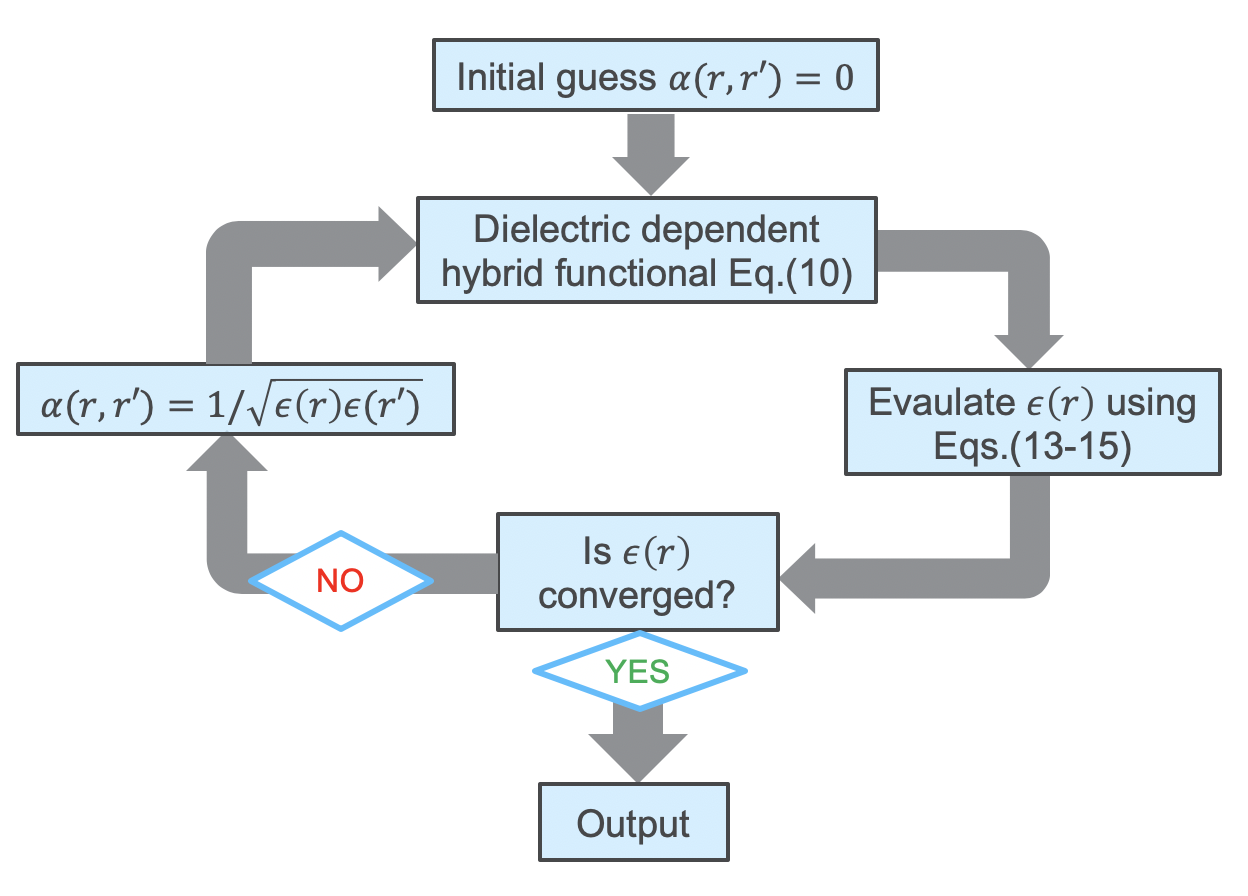}
    \caption{Dielectric dependent hybrid (DDH) functional calculations. In evaluating $\epsilon(\vec r)$, the derivatives entering Eq.~\eqref{eq:eps_ij} are computed numerically by performing two independent calculations with $E=\pm \delta$ a.u. and taking the difference, where $\delta$ is chosen small enough so as to insure calculations in the linear regime.}
    \label{fig:self_consistent}
\end{figure}

We computed the local dielectric function $\epsilon (\bf r)$ by minimizing the electric enthalpy [Eq.~\eqref{eq:E}] with the Kohn-Sham energy defined using the exchange correlation functional of Eq.~\eqref{eq:exc_hetero2}. The minimization is carried out using a finite field approach, as implemented in the Qbox code \cite{Gygi2008,Qboxurl}.
The function $\epsilon (\bf r)$ is computed  self-consistently. The whole procedure is schematically shown in Fig.~\ref{fig:self_consistent}.  At the first iteration we perform a DFT calculation at the PBE level [$\alpha(\vec r, \vec r^\prime)=0$]. At the second iteration we set $\epsilon(\vec r) = \epsilon^\text{PBE}(\vec r)$ in Eq.~\eqref{eq:separableform} and repeat the process until $\epsilon(\vec r)$ and the total energy are converged. 

\section{Validation of  self-consistent hybrid functionals for  3D \& 2D materials, surface \& interfaces}
\label{sec:validation}
\subsection{ Three-dimensional and two-dimensional materials}
\label{sec:bulk}
\begin{figure*}[hbt]
\centering
\includegraphics[width=0.90\linewidth]{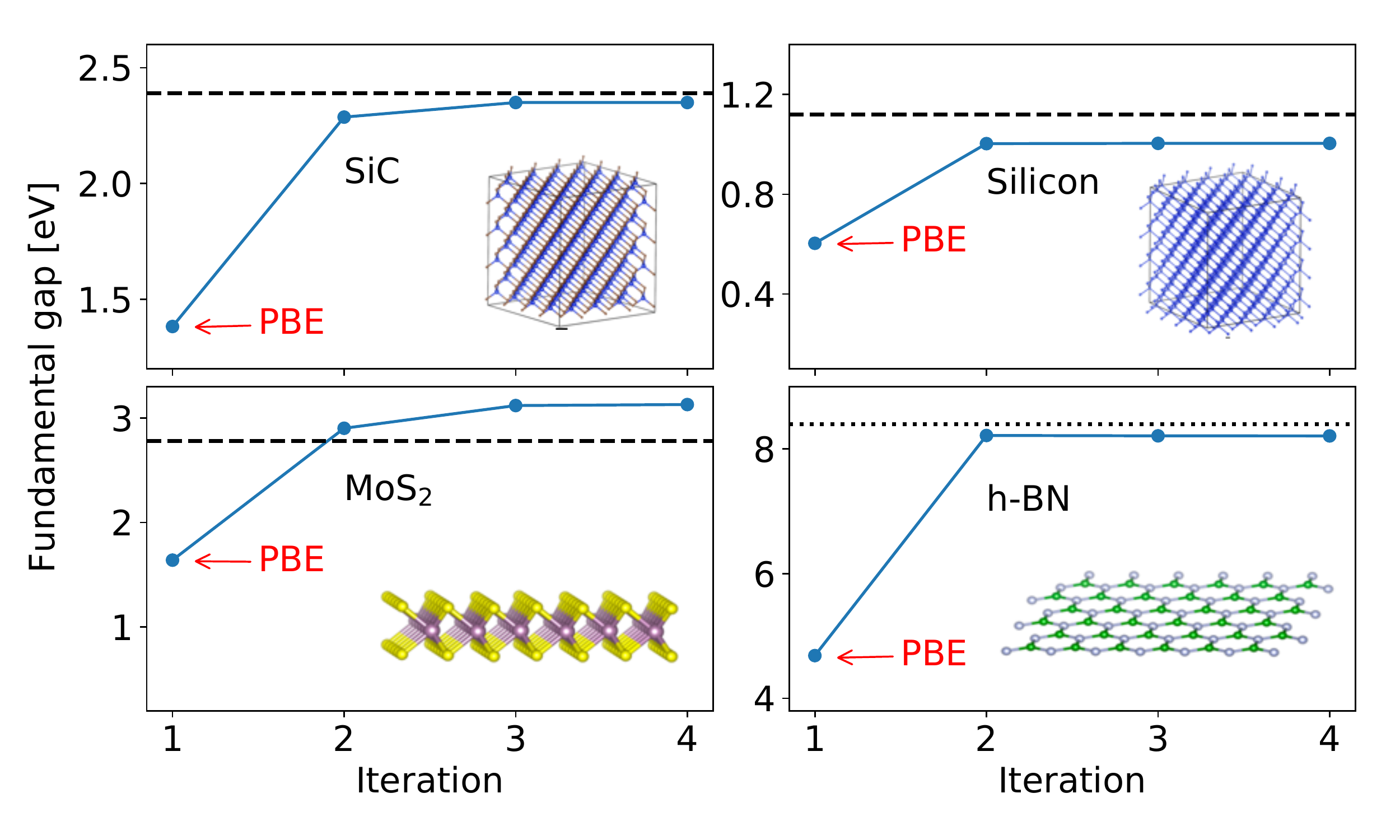}
\caption{Fundamental electronic gaps  of 3D solids, 3C-SiC and Si (upper panel), and 2D materials, MoS$_2$ and h-BN (lower panel), computed using the functional of Eq.~\eqref{eq:exc_hetero2}, as a function of the number of iteration of the self-consistent procedure (see Fig.~\ref{fig:self_consistent}). The horizontal dash lines denote experimental values. The dotted line (lower panel, right) is the self-consistent GW result for h-BN from Ref. \cite{Berseneva2013}.}\label{fig:schyb_gap}
\end{figure*}

Fig.~\ref{fig:schyb_gap} shows the band gap at each iteration for bulk Si and a 3C-(SiC) computed using  supercells with 512 atoms and the $\Gamma$ point to sample the Brillouin zone (the corresponding $\epsilon(z)$ [average of $\epsilon(\vec r)$ in the xy plane] are shown in the supplementary information [SI]).  In both cases, calculations rapidly converge and the computed band gap agrees with the experimental one within $\sim0.1$ eV (see Table.~\ref{tab:epsz_hmg}). The results for dielectric constants and band gaps of several solids, including covalently, ionic and van der Waals bonded systems,  are shown in Tables.~\ref{tab:epsz_hmg} and ~\ref{tab:gap_hmg}, respectively. Our results for the dielectric constants are all close to those of  self-consistent hybrid calculations reported in Ref.~\cite{Skone2014} [using the functional of Eq.~\eqref{eq:exc_hetero2} with  $\epsilon_{\infty}$ replacing $\epsilon(\vec r)$]. The use of the microscopically averaged $\epsilon$ over  the whole cell appear to yield results in slightly better agreement with experiments. Part of the small differences between column 3 and 4 in Table I is due to the use of pseudopotentials (this work) versus all electron calculations (Ref.~\cite{Skone2014}). 

Table II shows band gaps obtained with the functional of Eq.~\eqref{eq:exc_hetero2} and the procedure shown in Fig.~\ref{fig:self_consistent} (column 3) and those obtained with the global hybrid functional defined in Ref.~\cite{Skone2014} with two different values of $\epsilon_{\infty}$: the bulk average of $\epsilon (\vec r)$ computed in this work (column 4), and the $\epsilon_{\infty}$  from Ref.~\cite{Skone2014} (column 5). 
Considering that the all-electron results of Ref.~\cite{Skone2014} (reported in column 6) are obtained with all electrons and a localized basis set, the comparison between columns 5 and 6 shows differences arising from the use of pseudopotentials and the plane-wave basis set. The comparison between column 4 and 5 shows the sensitivity of the band gaps to slightly different values of $\alpha$. The most interesting comparison is between column 3 and 4 which shows that  the spatial variations of $\epsilon (\vec r)$ hardly affect the band gap of covalently bonded systems; however they do influence the computed gap for ionic and especially van der Waals bonded solids. 

\begin{table}[hbt]
\centering
\caption{The electronic dielectric constants ($\epsilon_\infty$) of three dimensional materials  obtained from PBE and spatial-resolved dielectric dependent hybrid functional (DDH) calculations [Eq.~\eqref{eq:exc_hetero2}], compared  with the results of  Ref.~\cite{Skone2014} and experiment. All calculations (PBE and DDH) were carried out using ONCV pseudopotentials~\cite{Schlipf2015} and by sampling the Brillouin zone with the ${\Gamma}$ point. The number of atoms or units used in the supercell calculations are indicated as subscripts for each solids. }\label{tab:epsz_hmg}
\begin{tabular}{c||c|c|c|c}
\hline
& PBE & DDH & Ref.~\cite{Skone2014} & Exp. \\
\hline
\hline
Si& 12.46 & 11.80  & 11.76 & 11.9 \cite{Yu_2010}  \\ 
SiC&6.86 & 6.49  & 6.50 & 6.52 \cite{Yu_2010} \\
AlP	& 8.08 & 7.57 & 7.23 & 7.54 \cite{Yu_2010}  \\
Diamond	& 5.77 & 5.58 & 5.61 &  5.70 \cite{Yu_2010} \\
MgO& 3.26 & 2.99 & 2.81  & 2.96 \cite{Noauthor_crc_2009}\\
LiCl &	2.93 & 2.77 & 2.77 & 2.70 \cite{Van_1969} \\
Ar & 	1.73 & 1.66 & 1.66 &  1.66 \cite{PhysRev.181.1297}\\
Ne &  1.29 & 1.25 & 1.21 & 1.23 \cite{schulze_density_1974} \\
\hline
\end{tabular}
\end{table}

\begin{table}[hbt]
\centering
\caption{The fundamental energy gaps (eV) of three dimensional materials  obtained from PBE and spatial-resolved dielectric dependent hybrid functional (DDH) calculations [Eq.~\eqref{eq:exc_hetero2}], compared with the results of Ref.~\cite{Skone2014} and experiment. All calculations (PBE and DDH) were carried out using ONCV pseudopotentials~\cite{Schlipf2015} and by sampling the Brillouin zone with the ${\Gamma}$ point. The number of atoms or units used in the supercell calculations are indicated as subscripts for each solids. In columns 4 and 5 we report calculations with a constant mixing fraction (See Eq.~\eqref{eq:exc_hetero1}),  $\alpha=1/\bar\epsilon$, and $\alpha=1/\epsilon_\infty$ respectively. The zero-phonon renormalization (ZPR) is reported when available from experiment~\cite{Chen2018}.}\label{tab:gap_hmg}

\begin{tabular}{c||c|c|c|c|c|c|c}
\hline
& PBE & DDH&$\alpha = 1/{\bar \epsilon}$ \footnote{Hybrid functional calculation with $\alpha=1/{\bar \epsilon}$ where $\bar \epsilon$ is the bulk average of $\epsilon(\vec r)$: values reported in Table~\ref{tab:epsz_hmg}.}& $\alpha=1/\epsilon_\infty$\footnote{Hybrid functional calculation with $\alpha=1/\epsilon_\infty$ where $\epsilon_\infty$ is  from Ref.~\cite{Skone2014}: values reported in Table~\ref{tab:epsz_hmg}.}& Ref.~\cite{Skone2014} & ZPR & Exp.\\
\hline
\hline
Si & 0.603 & 1.00  & 1.01 & 1.01 & 0.99 & 0.06 & 1.17 \cite{kittel_introduction_2004}\\ 
SiC& 1.38 & 2.35 & 2.35 & 2.35 &  2.29 & 0.11 & 2.39 \cite{PhysRev.133.A1163} \\
AlP & 1.56 & 2.27& 2.28 & 2.32 & 2.37  &  0.02 & 2.51\cite{PhysRevB.8.5711}  \\
Diamond	& 4.17 &  5.48 &  5.54 &5.53 &5.42  & 0.37 & 5.48 \cite{clark_c._d._intrinsic_1964}\\ 
MgO& 4.78 & 7.70 & 8.08 & 8.30 & 8.33 & 0.53 & 7.83 \cite{whited_exciton_1973}\\
LiCl & 6.47 & 9.38& 9.56 & 9.56 & 9.62 & 0.17 &  9.40 \cite{baldini_optical_1970} \\
Ar &  8.70 & 13.93 & 14.34 & 14.34 & 14.67 & & 14.2 \cite{PhysRevLett.34.528} \\
Ne & 11.62 & 20.60& 22.38 & 22.72 & 23.67 &   & 21.7 \cite{PhysRevLett.34.528}\\
\hline
\end{tabular}
\end{table}

Fig.~\ref{fig:schyb_gap} shows the band gap for monolayer MoS$_2$ and h-BN.  The dielectric hybrid  hybrid functional (DDH) of Eq.~\eqref{eq:exc_hetero2} predicts a fundamental gap of 3.1 eV for MoS$_2$. The effect of spin-orbit coupling, known to lead to a splitting of the degenerate valence bands of about 0.1 eV~\cite{Cheiwchanchamnangij2012}, was neglected in our calculations. Therefore, we conclude that our quasiparticle gap is in reasonable agreement with the experimental value of 2.78(2) eV~\cite{Yao2017}.
 
The self-consistent hybrid functional of Eq.~\eqref{eq:exc_hetero2} predicts a  gap of 8.2 eV for h-BN. This is consistent with that obtained with  self-consistent GW calculations  ($\sim 8.4$ eV) in Ref.~\cite{Berseneva2013}. The Kohn-Sham gap obtained in PBE calculations is about 4 eV smaller, and G$_0$W$_0$ and GW$_0$ results  using PBE wavefunctions also underestimate the quasiparticle gap by $\sim 2$ and $\sim 1$ eV, respectively \cite{Berseneva2013,Smart2018}. 

The dielectric function $\epsilon(z)$ of the 2D systems studied here turns out to be localized at the  monolayers (see Fig.~S3 in Ref.~\cite{SM}). This provides a physical measure  of the ``dielectric thickness" of the 2D layers, which we define as $
w_\epsilon = \frac{\int dz(z-z_0)^2 \chi(z)}{\int dz \chi(z)}$
where $\chi(z) := \epsilon(z) - 1\,$. We obtain a thickness of 3.4 and 1.6 Bohr for MoS$_2$ and h-BN respectively. The spreads of the charge density (see Fig.~S3 in Ref.~\cite{SM}) are  2.3 and 1.3 Bohr respectively, slightly smaller than those of the respective dielectric functions, but comparable.  

\subsection{Surfaces and interfaces}
\label{sec:surfaces_interfaces}
In the case of surfaces and interfaces,  we carried out calculations with the scheme outlined in Fig.~\ref{fig:self_consistent}, applying the $\vec E$ field parallel to the surface/interface, insuring that the tangential part of the $\vec E$ field is  continuous across the interface. (If a constant $\vec D$ field were applied, when minimizing the functional Eq.~\eqref{eq:D_functional}, the $\vec D$ field would be instead perpendicular to the interface). 

\begin{figure}[hbt]
\centering
\includegraphics[width=0.45\textwidth]{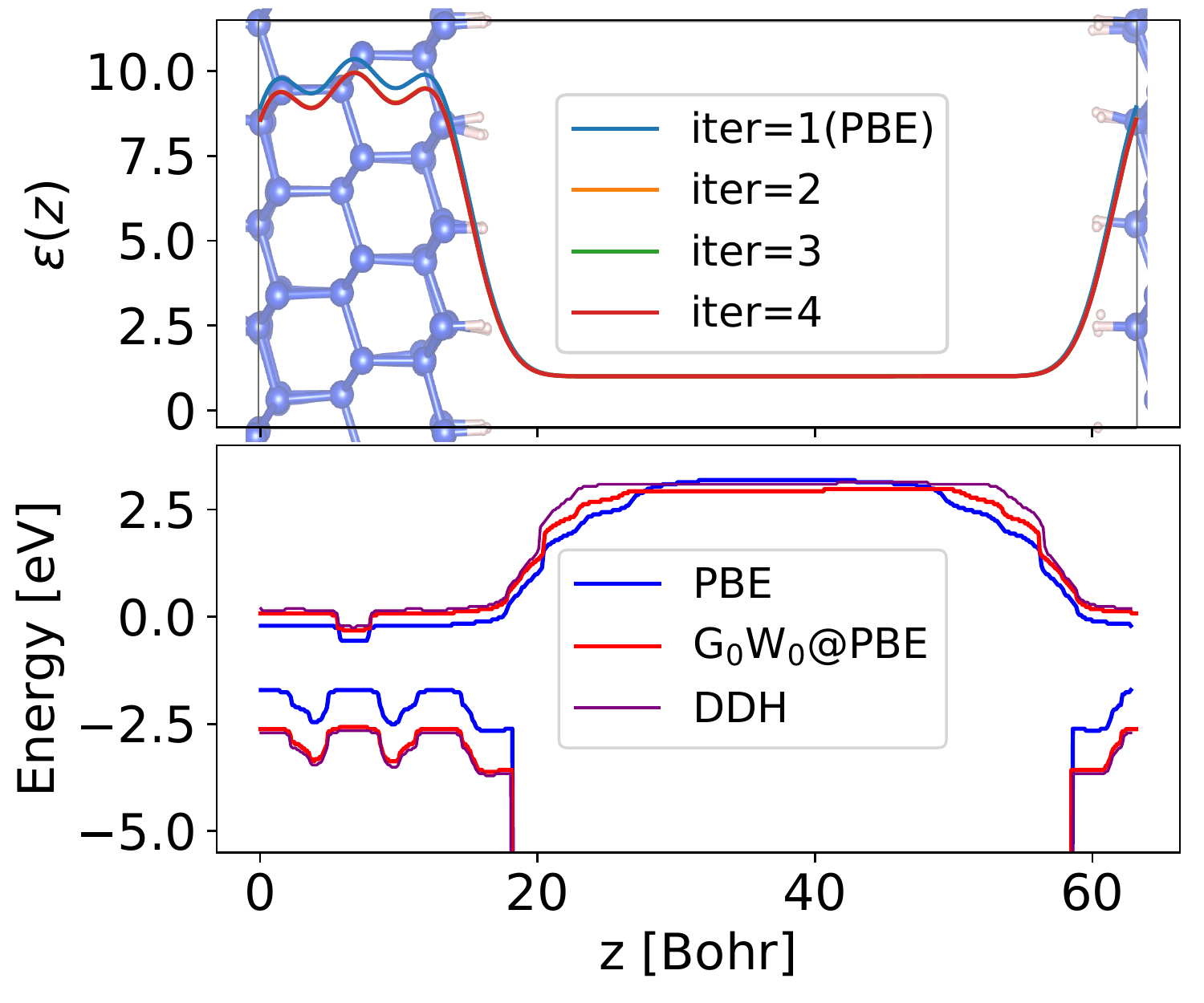}
\caption{The local dielectric function $\epsilon (z)$ of the unreconstructed, H-terminated Si(111) surface (Si-H),  (average of  $\epsilon (\vec r )$ over the (x,y) plane) is plotted as a function of z, the direction perpendicular to the surface, on the upper panel.  We show values obtained as a function of the number of iterations, when using the procedure outlined in Fig.5.  The band offsets between  the H-Si surface and vacuum, computed  at different levels  of theory, are shown on the right panel.  We show results computed with the functional of Eq.~\eqref{eq:exc_hetero1}, PBE and the G$_0$W$_0$@PBE level of theory, obtained with the WEST code.}\label{fig:bo_surface}
\end{figure}

Fig.~\ref{fig:bo_surface} shows the dielectric function and band offsets for an unreconstructed, hydrogen terminated silicon (111) surface (H-Si). We find that the dielectric constant in the silicon bulk regions is $\sim$ 9, which is  smaller than that reported in Table~\ref{tab:epsz_hmg}, due to finite size effects. Indeed, the silicon slab has only 72 Si atoms, a size insufficient to converge the dielectric constant to the bulk value. 

The band gap of the silicon portion of the slab and the band offsets between the surface and vacuum obtained from  DDH  calculations are in good agreement with those of  G$_0$W$_0$@PBE calculations; we note  that there is a slight difference in the spatial variation of the conduction band  at the interface, which is sharper in the case of the hybrid functional calculations, possibly indicating differences between the PBE wavefunctions and charge density (not updated in the GW calculations)  and the respective quantities computed self-consistently at the hybrid level. 

\begin{figure*}[hbt]
\centering
\includegraphics[width=0.90\textwidth]{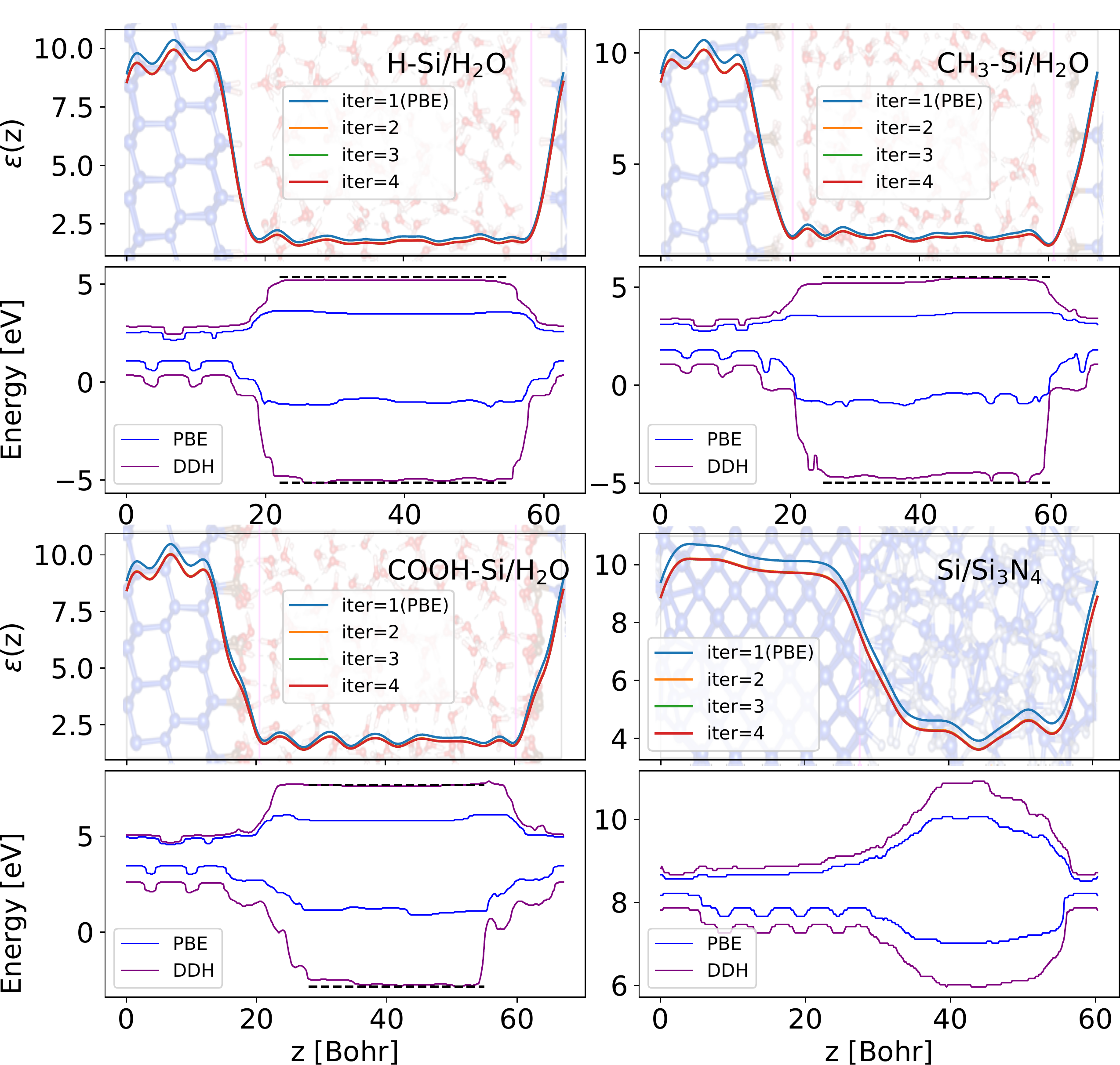}
\caption{Dielectric function [$\epsilon(z)$, average of $\epsilon (\vec r)$ in the (x,y) plane] and band offsets of four interfaces computed using the DDH  functional of Eq.~\eqref{eq:exc_hetero1}. The dielectric function $\epsilon(z)$ is  computed using the method outlined in Fig.~\ref{fig:self_consistent}; results are shown as a function of the number of iterations. The direction $z$ is perpendicular to the interface. The electric field is applied along the $x$ direction. The dashed lines for the band offsets of aqueous interfaces are the results of G$_0$W$_0$@DDH calculations of water from Ref.~\cite{Gaiduk2018}, with the conduction band of H$_2$O aligned with the minimum of the conduction band of the corresponding interface.}\label{fig:epsz_interface}
\end{figure*}

Calculations for representative interfaces (H-Si/H$_2$O, CH$_3$-Si/H$_2$O, COOH-Si/H$_2$O, and Si/Si$_3$N$_4$) are shown in Fig.~\ref{fig:epsz_interface}.
We again observe that the calculation of $\epsilon(\vec r)$ converges rapidly, after 3-4 iterations (see Fig.~\ref{fig:epsz_interface}, upper panels). 
We can clearly see that there are two distinct average values of $\epsilon$ in the two bulk regions where $\epsilon$ oscillates around a constant value. The transition regions in the four interfaces, defined as the region where $\epsilon(\vec r)$ changes sharply, have a thickness of approximately 5 Bohr for aqueous interfaces and 10 Bohr for the silicon-silicon nitride interface.

As already found for the hydrogenated Si-surface, the DDH functional of Eq.~\eqref{eq:exc_hetero2} predicts the band gap in the silicon bulk regions (Fig.~\ref{fig:epsz_interface}) in agreement with G$_0$W$_0$@PBE. In the water region of the aqueous interfaces however, the VBM and CBM are substantially different from those predicted by  G$_0$W$_0$@PBE calculations; this is understandable since the PBE wavefunctions are not a good approximation of the band edges of water, as shown in Ref.~\cite{Gaiduk2018}. The DDH calculations are instead in good agreement with the values reported in Ref.~\cite{Gaiduk2018} and obtained at the G$_0$W$_0$@sc-hybrid level, where the mixing fraction was taken equal to the electronic dielectric constant of water. The band gap ($10.5$ eV) is also in good agreement with that found in Ref.~\cite{Gaiduk2018}. In the case of  Si/Si$_3$N$_4$ (Table~\ref{tab:sisi3n4})  we compare our DDH results with experiment,  and we find good agreement (the band gap of silicon is again larger than in experiment, due to finite size effects, i.e. to the small slab chosen in our calculations). 
\begin{table}[hbt]
\caption{Band offsets (eV)  computed at different levels of theory (using the PBE functional and the functional of Eq.~\eqref{eq:exc_hetero2}, with the procedure of Fig.~\ref{fig:self_consistent}) for the silicon-silicon nitride interface,  compared with experiment (from \cite{Anh2013} and the references therein)}\label{tab:sisi3n4}
\begin{tabular}{c||c|c|c}
\hline
 Si/Si$_3$N$_4$& PBE  & DDH & Exp.  \\
\hline
\hline
Conduction band offset & 1.2  & 1.9 & 1.83 $-$ 2.83 \\
Valence band offset  & 0.7 & 1.3 & 1.5 $-$ 1.78\\
\hline
\end{tabular}
\end{table}

\section{Conclusions}
\label{sec:conclusion}
We introduced a general dielectric-dependent functional, which is applicable to any semiconductor and insulator and does not contain any adjustable parameter. The functional is a generalization of the self-consistent hybrid functional for homogeneous solids introduced in Ref.~\cite{Skone2014}, and it is defined using a local, spatially dependent dielectric function. We justified the definition of the functional and the spatial variation of the dielectric function using the disentanglement of the dielectric spectra of heterogeneous systems in terms of  the spectra of subsystems; such a disentanglement was achieved  using linear combinations of dielectric eigenvectors localized in real space. The local dielectric function was then computed self-consistently by carrying our density functional calculations in finite electric fields.

We showed that the dielectric hybrid functional introduced here  predicts the band gaps and dielectric constants of three- and two-dimensional solids, as well as band offsets of surfaces and interfaces, with an accuracy  comparable to that of GW calculations, thus paving the way to efficient and accurate calculations of the electronic properties of complex heterogeneous systems. 

Finally we note that the formulation introduced in our work provides a  definition of the dielectric thickness  of interfaces and 2D systems, and a physical interpretation of the spatial variations of single particle energy levels upon the formation of interfaces.

\begin{acknowledgments}
We thank Francois Gygi, Christopher Knight, and Jonathan Skone for numerous discussions. This work was supported by Argonne Leadership Computing Facility (ALCF) Theta Early Science program and the Midwest Integrated Center for Computational Materials (MICCoM). This work used computing resources of Argonne Leadership Computing Facility, which is a DOE Office of Science User Facility under Contract DE-AC02-06CH11357. We also gratefully acknowledge the computing resources provided on Bebop, a high-performance computing cluster operated by the Laboratory Computing Resource Center at Argonne National Laboratory.
\end{acknowledgments}
\bibliographystyle{apsrev4-1prx}
\bibliography{hybrid}
\end{document}